\documentclass[conference]{IEEEtran}
\IEEEoverridecommandlockouts
\usepackage{cite}
\usepackage{amsmath,amssymb,amsfonts}
\usepackage{amsthm}  % ???? theorem-like ??
\usepackage{subfigure}%??
\usepackage{algorithm} 
\usepackage{algpseudocode}
\usepackage{graphicx}
\usepackage{textcomp}
\usepackage{xcolor}
\usepackage{booktabs}
\usepackage{multirow}%????
\def\BibTeX{{\rm B\kern-.05em{\sc i\kern-.025em b}\kern-.08em
    T\kern-.1667em\lower.7ex\hbox{E}\kern-.125emX}}
\usepackage{mathptmx}
\newtheorem{definition}{Definition}

\begin{document}

\title{VariSAC: V2X Assured Connectivity in RIS-Aided ISAC via GNN-Augmented Reinforcement Learning}%\\
%\title{HierRole: Hierarchical Structure Learning in Hyperbolic Space for Role Discovery in Online Social Networks\\
%HierRole: Hierarchical Structure Learning based Role Discovery in Online Social Networks
%HierRole: A Simplified Hyperbolic Graph Transformer for Role Discovery in Online Social Networks

%\maketitle
%\thanks{This work is supported by the Zhejiang Provincial Natural Science Foundation of China under Grant LDT23F01012F01 and by the National Natural Science Foundation of China under Grant Number 62401190, 62372146 and 62071327. Corresponding author: Pengfei Jiao.}}

\author{\IEEEauthorblockN{Huijun Tang\IEEEauthorrefmark{1}, Wang Zeng\IEEEauthorrefmark{2}, Ming Du\IEEEauthorrefmark{2}, Pinlong Zhao\IEEEauthorrefmark{2}, Pengfei Jiao\IEEEauthorrefmark{2}, Huaming Wu\IEEEauthorrefmark{3} and Hongjian Sun\IEEEauthorrefmark{1}}
\IEEEauthorblockA{\IEEEauthorrefmark{1}Department of Engineering, Durham University, Durham DH1 3LE, United Kingdom}
\IEEEauthorblockA{\IEEEauthorrefmark{2}School of Cyberspace, Hangzhou Dianzi University, Hangzhou 310018, China}
\IEEEauthorblockA{\IEEEauthorrefmark{3}Center for Applied Mathematics, Tianjin University, Tianjin 300072, China}
Emails: huijun.tang@durham.ac.uk, \{242270069, mdu, pinlongzhao, pjiao\}@hdu.edu.cn, \\whming@tju.edu.cn, hongjian.sun@durham.ac.uk}
% \and
% \and
% \and
% \and
% \and

\maketitle

\begin{abstract}
The integration of Reconfigurable Intelligent Surfaces (RIS) and Integrated Sensing and Communication (ISAC) in vehicular networks enables dynamic spatial resource management and real-time adaptation to environmental changes. However, the coexistence of distinct vehicle-to-infrastructure (V2I) and vehicle-to-vehicle (V2V) connectivity requirements, together with highly dynamic and heterogeneous network topologies, presents significant challenges for unified reliability modeling and resource optimization. To address these issues, we propose VariSAC, a graph neural network (GNN)-augmented deep reinforcement learning framework for assured, time-continuous connectivity in RIS-assisted, ISAC-enabled vehicle-to-everything (V2X) systems. Specifically, we introduce the Continuous Connectivity Ratio (CCR), a unified metric that characterizes the sustained temporal reliability of V2I connections and the probabilistic delivery guarantees of V2V links, thus unifying their continuous reliability semantics. Next, we employ a GNN with residual adapters to encode complex, high-dimensional system states, capturing spatial dependencies among vehicles, base stations (BS), and RIS nodes. These representations are then processed by a Soft Actor-Critic (SAC) agent, which jointly optimizes channel allocation, power control, and RIS configurations to maximize CCR-driven long-term rewards. Extensive experiments on real-world urban datasets demonstrate that VariSAC consistently outperforms existing baselines in terms of continuous V2I ISAC connectivity and V2V delivery reliability, enabling persistent connectivity in highly dynamic vehicular environments.
%The integration of Reconfigurable Intelligent Surfaces (RIS) and Integrated Sensing and Communication (ISAC) in vehicular networks enables efficient spatial resource reuse and real-time adaptation to environmental changes. However, high mobility and frequent obstructions lead to rapidly evolving network topologies, which require advanced state representations to capture complex spatio-temporal interactions among vehicles, roadside units (RSUs), and infrastructure for reliable service continuity. To address these challenges, we propose VariSAC, a graph neural network (GNN)-augmented deep reinforcement learning framework for assured, time-continuous connectivity in RIS-assisted ISAC-enabled vehicle-to-everything (V2X) networks. VariSAC integrates RIS to maintain virtual line-of-sight links and employs ISAC to jointly optimize communication and sensing resources. A sliding time window and temporally-aware reward function are introduced to enforce service continuity across consecutive time steps. To model the dynamic V2X topology, a GNN with a residual adapter is utilized for global state encoding, effectively capturing intricate spatial and temporal dependencies among vehicles, RSUs, and RIS elements. Extensive evaluations on real-world urban roadmaps and vehicle trajectory datasets show that VariSAC outperforms state-of-the-art baselines in vehicle-to-infrastructure (V2I) throughput, vehicle-to-vehicle (V2V) reliability, and task completion rate, enabling robust and persistent connectivity in dynamic vehicular environments.%??
\end{abstract}


\section{Introduction}
The rapid advancement of wireless communication technology has become a critical enabler for autonomous vehicles and intelligent transportation systems. However, it also imposes increasingly stringent requirements on network performance and reliability. %but it also imposes more stringent requirements on network performance and reliability. 
Integrated Sensing and Communication (ISAC) has emerged as a promising paradigm that unifies sensing and communication functionalities through shared spectrum and hardware resources. This integration enables vehicular networks to simultaneously support both high-rate data transmission and real-time environmental awareness~\cite{10944644,11062661,9737357}. By leveraging joint spectrum and hardware usage, ISAC significantly enhances spectral efficiency~\cite{9779322}.

%Integrated Sensing and Communication (ISAC) enables vehicular systems to simultaneously support high-rate data transmission and environmental sensing by sharing spectrum and hardware resources, significantly improving spectral efficiency~\cite{10944644,11062661,9779322}. 
While significantly enhancing the performance of Vehicle-to-Everything (V2X) systems, ISAC technology introduces new challenges for resource allocation~\cite{twardokus2022vehicle,wei2023integrated}. The deep integration increases spectral efficiency, but also complicates system-wide resource management due to the need for joint scheduling of sensing and communication resources in real time~\cite{ma2020joint,liu2020joint}. For example, high-precision environmental perception demands stable and sufficient bandwidth allocation to support centimeter-level positioning and low-latency obstacle detection~\cite{caillot2022survey}. Moreover, the performance of ISAC-based V2X is highly dependent on the availability of line-of-sight (LoS) links, with significant degradation in the presence of buildings, vegetation, or other obstacles~\cite{sun2024performance}. To address these issues, spatial resource reuse?where signals are multiplexed in space via beamforming or intelligent reflection?has emerged as a key enabler for next-generation V2X networks~\cite{8354811}. Reconfigurable Intelligent Surfaces (RIS) offer a programmable and cost-effective solution by establishing virtual LoS links and dynamically manipulating wireless propagation~\cite{10876793,wu2019towards,9133130}.

By adaptively adjusting the phase of incident signals, RIS can create alternative paths, enable spatial separation of simultaneous transmissions, and help maintain robust connectivity and sensing performance?even in complex, rapidly changing urban environments~\cite{9122596,shah2023effective}. When combined with ISAC, however, maintaining connectivity becomes even more complex, as both communication and tightly coupled sensing tasks must be jointly supported. In highly dynamic vehicular environments, frequent link quality fluctuations or intermittent degradation can impair collaborative awareness, reduce service availability, and jeopardize safety-critical functions~\cite{9382930,liu2022integrated}. These challenges are further compounded by the demands of advanced applications such as autonomous driving and cooperative perception, which require stable end-to-end links despite high mobility, rapidly changing topologies, frequent non-line-of-sight (NLoS) conditions, and varying interference levels in urban scenarios.
To address these obstacles, various connectivity maintenance strategies have been proposed, including diversity-based methods, handover optimization, and adaptive resource allocation~\cite{souri2024systematic,10726620,9521309}. Nevertheless, most existing approaches primarily focus on physical-layer reliability or instantaneous connectivity, and often fall short in capturing the pronounced spatio-temporal dynamics and the joint communication-sensing requirements that characterize RIS-assisted ISAC-enabled V2X systems. 

Despite the potential of RIS-assisted ISAC for V2X, several critical challenges remain:
\begin{itemize}
\item \textbf{Divergent Continuous Connectivity Requirements Across V2X Links:} V2I links, supporting both sensing and infotainment, require sustained link quality above distinct sensing and communication thresholds to ensure continuous perception and stable data streaming. In contrast, V2V links prioritize the timely and reliable delivery of safety-critical messages within strict latency bounds. These distinct connectivity semantics (continuous versus one-shot) pose key challenges for unified reliability modeling and resource optimization. 
%These fundamentally different connectivity semantics?continuous duration versus one-shot delivery?pose significant challenges for unified reliability modeling and resource optimization. 
The presence of RIS further complicates this issue by dynamically altering the propagation environment, inducing link fluctuations that simultaneously affect sensing and communication, and making it more difficult to guarantee differentiated connectivity requirements.
%The introduction of RIS further exacerbates this issue by dynamically modifying the propagation environment, causing link fluctuations that jointly impact sensing and communication, and complicating the assurance of differentiated connectivity.

  %These distinct requirements arise from the nature of the services they support?continuous and bandwidth-intensive for V2I, versus sporadic and delay-sensitive for V2V. As such, a unified connectivity framework must accommodate both the temporal continuity of V2I links and the probabilistic transmission success of V2V links, which cannot be jointly captured by conventional reliability metrics.
  %  \item \textbf{Continuous Connectivity Guarantee:} Ensuring that each vehicle maintains reliable connectivity over extended periods---rather than at isolated time points---is essential for safety-critical applications, yet remains largely unaddressed.
%    \item \textbf{Joint Resource Allocation under Coupled Sensing and Communication:} The tight coupling of sensing and communication tasks in ISAC makes real-time, fine-grained resource allocation significantly more complex, especially under dynamic and heterogeneous conditions.
 %   \item \textbf{Capturing Complex Network Dynamics:} Rapid changes in network topology and strong spatial-temporal correlations challenge existing methods, which often fail to represent the deep relationships between vehicles, infrastructure, and RIS nodes.
 \item \textbf{Dynamic and Heterogeneous V2X Network Topologies:} RIS-assisted ISAC-enabled V2X systems operate in highly dynamic environments, where vehicle mobility, intermittent LoS conditions, and frequent RIS reconfigurations introduce strong spatio-temporal dependencies across links. Meanwhile, V2X entities, including vehicles, RSUs, and RIS nodes, exhibit heterogeneous capabilities in sensing, communication, and control. The resulting diversity in link types (e.g., V2V, V2I, V2RIS) and service roles further complicates unified modeling. These dynamic and structural complexities are difficult to capture with conventional approaches.
\end{itemize}

To tackle the above challenges, we present \textbf{VariSAC}, a novel framework for assured V2X connectivity in RIS-assisted ISAC-enabled networks, empowered by GNN-augmented reinforcement learning. VariSAC introduces a continuous connectivity ratio for individual vehicles, based on multi-slot temporal assessment, enabling consistent maintenance of link reliability over time for both V2I and V2V services. The main contributions of this paper are summarized as follows:
\begin{itemize}
   % \item To address the problem that existing vehicular network resource allocation schemes neglect the service quality guarantee for individual vehicles, this paper proposes a continuous connectivity metric modeling method. By introducing a reliable connection measurement mechanism based on multi-slot continuous evaluation, the method can effectively quantify the connection stability of individual vehicles in high-mobility environments.
    %????????????????????????????????????????????????????????????????????????????????????????????
    %Thj: ????????????Reliable Connectivity
\item \textbf{CCR-Guided Joint Optimization with Sliding Window Reliability Modeling:} 
To address the divergent reliability semantics of V2I and V2V services, we introduce a unified metric termed Continuous Connectivity Ratio (CCR), which captures both the duration-based SNR requirements of V2I sensing/communication and the probabilistic delivery constraints of V2V safety messaging. Guided by CCR, we design a joint optimization framework that leverages a sliding time window to explicitly model temporal continuity, and jointly optimizes channel resource allocation, power control, and RIS phase-shift configurations to maintain continuous reliability.
\item \textbf{GNNRA-Augmented Reinforcement Learning for Continuous Connectivity:}
We propose a novel integration of graph neural networks with residual adapters (GNNRA) and deep reinforcement learning (DRL) to address the continuous reliable connectivity of RIS-assisted ISAC-enabled vehicular networks. The GNNRA module captures the heterogeneous and dynamic spatial relationships among vehicles, RSUs, and RIS nodes. These enriched spatial embeddings are fed into a DRL agent, which performs long-term connectivity optimization under communication and sensing constraints, thereby enabling divergent V2X services with continuous QoS assurance in highly dynamic environments.

    \item \textbf{Comprehensive Evaluation on Real-World Urban Topologies:}
We conduct extensive experiments using both large-scale simulation data and real-world urban roadmaps with actual vehicle trajectory datasets. The results demonstrate the effectiveness of the proposed method.

 %   \item A dual validation system based on simulation data and real vehicle trajectories is established. Through comparative analysis of uniform simulation scenarios and real trajectory scenarios with uneven spatiotemporal distribution, the effectiveness and robustness of the proposed method are comprehensively validated in both idealized environments and complex real-world environments.
    %?????????????????????????????????????????????????????????????????????????????????????
    %Thj???????????????????
\end{itemize}

\section{System Model and Problem Formulation}
%\subsection{Considered Scenario}

\begin{figure}[ht!]
    \centering
    \vspace{-0.1in}
    \includegraphics[width=1\linewidth]{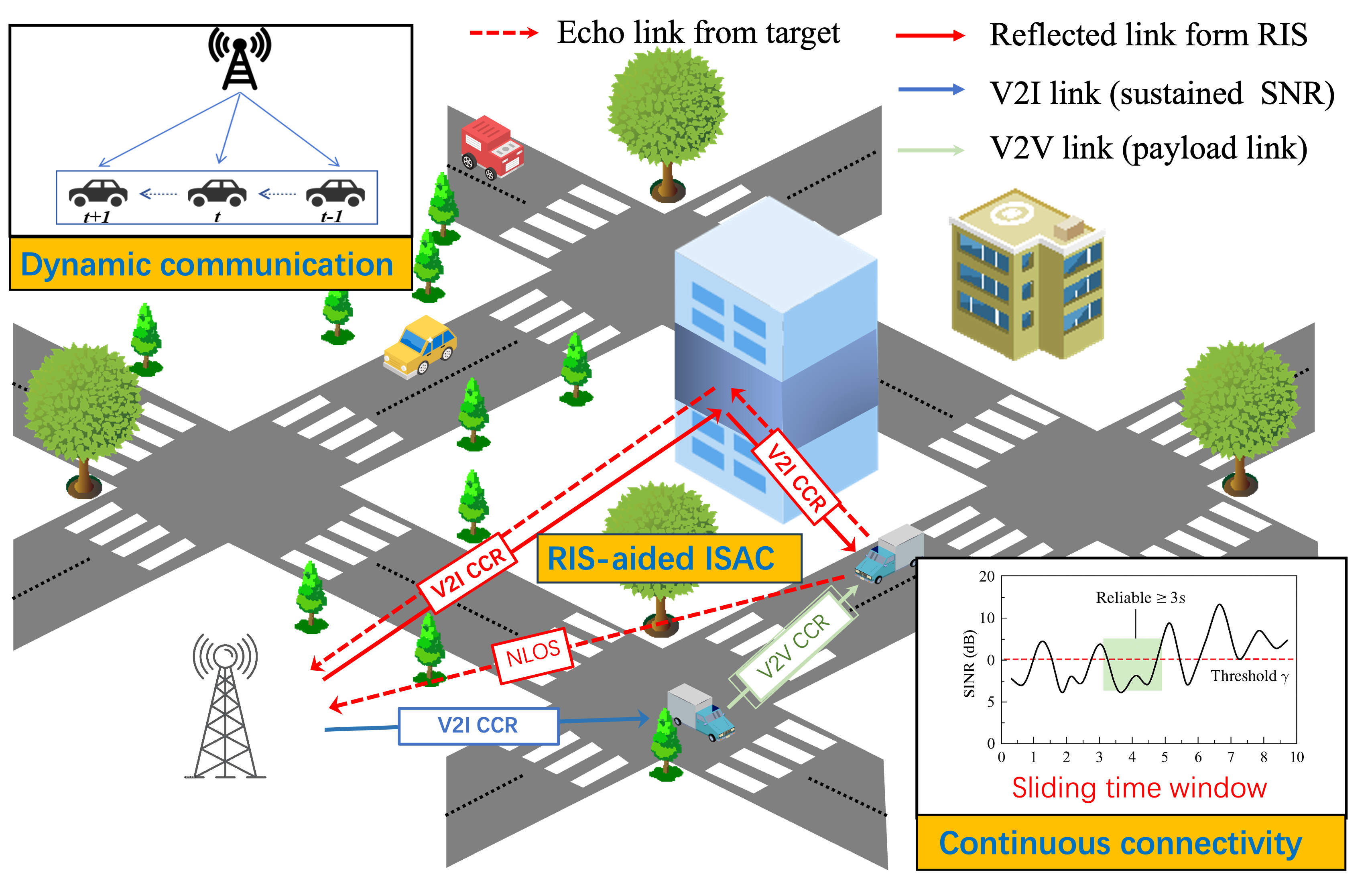}
   \vspace{-0.2in}
   \caption{System Model. RIS provides virtual LoS paths under blockage. V2I links require sustained SNR for sensing and communication (V2I CCR), while V2V links ensure timely payload delivery under latency bounds (V2V CCR).}
    \label{fig:enter-label}
\end{figure}

Fig.~\ref{fig:enter-label} depicts a RIS-assisted ISAC system tailored for V2X communications. The system comprises three main components: i) a multi-antenna BS, ii) a RIS equipped with $F$ passive reflecting elements, and iii) $V$ single-antenna vehicles, among which $J$ vehicles are designated as sensing targets. The system establishes three categories of links: 
\begin{itemize}
    \item \textbf{V2I links}: $V$ uplinks from vehicles to the BS, each assigned a dedicated subchannel.

\item \textbf{V2V links}: $D$ direct links between vehicles, which reuse the V2I subchannels.

\item \textbf{Sensing links}: $J$ dedicated subchannels for vehicle detection, enabling the BS to sense target vehicles via reflected signals.
\end{itemize}
The available spectrum is divided into orthogonal subchannels, with $V$ subchannels assigned to V2I communication and $J$ to sensing. A binary scheduling variable $c_{v,d}^t$ indicates whether the $d$-th V2V link reuses the $v$-th V2I subchannel at time slot $t$ ($c_{v,d}^t=1$) or not ($c_{v,d}^t=0$).

%The system supports $V$ V2I links for communication between vehicles and base stations, $D$ V2V links for communication between vehicles, and $J$ dedicated sensing links for target vehicle detection. The system divides the available spectrum into multiple orthogonal subchannels, allocating $V$ subchannels to V2I links and $J$ subchannels to sensing links. The V2V links reuse the V2I subchannels, where a binary indicator $x_{v,d}^{t}$ determines whether the $d$-th V2V link at time slot $t$ shares the subchannel of the $v$-th V2I link $(c_{v,d}^{t} = 1)$ or not $(c_{v,d}^{t} = 0)$.
%dm: ??????
%Thj???????????????connectivity?reliable
%??????
%????????
%Reflected link??RIS?????

%The RIS optimizes communication quality and perception performance by dynamically adjusting the propagation direction of the wireless network. For each V2I link, the received signal at a vehicle comprises two components: a direct path from the BS, and a reflected path via the RIS. 
The RIS improves both communication and sensing by dynamically controlling wireless propagation. For each V2I link, the vehicle receives signals via a direct path from the BS and a reflected path through the RIS. 
%In the case of V2I links, each vehicle receives signals containing two propagation paths: one is the direct link from the base station to the vehicle, and the other is the enhanced link reflected by RIS. 
The RIS employs a uniform planar array of $F$ low-cost passive elements, each of which independently adjusts the phase and amplitude of incident signals. The reflection coefficient matrix at time $t$ is:  
\begin{equation}
    \Theta_t = \operatorname{diag}\left[\beta_1 e^{j\theta_1^t}, \beta_2 e^{j\theta_2^t}, \ldots, \beta_f e^{j\theta_f^t}\right], \forall f \in \{1,2,\cdots,F\},
\end{equation}
where $\theta_f^t$ denotes the phase shift of the $f$-th element, and $\beta_f$ represents its amplitude coefficient. Due to hardware limitations, the phase shifts are restricted to a discrete set of values: %Hardware constraints limit the phase shifts to discrete values: 
$\theta_f^t \in \left\{0, \frac{2\pi}{Q}, \ldots, \frac{2\pi(Q-1)}{Q}\right\}$, where $Q$ denotes the phase shift quantization level, controlling beamforming precision, while $\beta_f \in [0,1]$ adjusts reflection strength. This discrete configuration enables flexible beamforming while maintaining practical feasibility.
%dm: ??????

\subsection{Channel Model}    
%????
We assume the coordinates of the BS and RIS are $(X_B, Y_B, Z_B)$ and $(X_R, Y_R, Z_R)$, respectively. At time $t$, the position of each vehicle is $(X_{t}, Y_{t}, Z_{t})$. In the RIS-assisted ISAC system, the channel considers both the direct channel and the indirect channel, where the channel gain of the direct channel is expressed as~\cite{8723178}:
%???????RIS??????$?X_B, Y_B, Z_B?$?$?X_R, Y_R, Z_R?$$????t??????????X_{t}, Y_{t}, Z_{t}).??RIS???ISAC?????????????????????????????????
\begin{equation}
    g_{Tra,Rec}^{t} = \sqrt{\rho \beta \left(d_{Tra,Rec}^{t}\right)^{\eta}} \xi_{Tra,Rec}^{t},
\end{equation}
where the transmitter $Tra \in \{v,d,j,B\}$ and receiver $Rec \in \{v,d,j,B\}$ denote vehicles, dedicated sensing targets, or the BS. When $Tra$ and $Rec$ refer to the same vehicle or BS, the channel gain $g_{Tra,Rec}^{t}\triangleq 0$. $\rho$ is the path loss constant at the reference distance $d_{0}=1m$, $\beta $ is the log-normal shadowing random variable, $d_{Tra,Rec}^{t}$ represents the geometric distance between the transmitter and receiver in the $t$-th time slot, and $\eta$ is the corresponding path loss exponent. $\xi_{Tra,Rec}^{t}$ is the frequency-dependent small-scale fading power component between the transmitting and receiving ends at time $t$, assumed to follow an exponential distribution with unit mean. %Assuming all communication and sensing occur in the $t$-th time slot, the sensing channel gain from the BS to the $j$-th target vehicle is $g_{B,j}^{t}$. The channel gain between the vehicle and BS for the $v$-th V2I link is $g_{v,B}^{t}$. The channel gain from the $v$-th V2I user to the receiver of the $d$-th V2V link is $g_{v,d}^{t}$, where if $v$ and $d$ correspond to the same vehicle, then $g_{v,d}^{t} \triangleq 0$. The channel gain from the transmitter of the $d$-th V2V link to the BS is $g_{d,B}^{t}$, and the channel gain to the receiver of the $d^{'}$-th V2V link is $g_{d,d^{'}}^{t}$.
%???$\rho$?????$d_{0}=1m$?????????$?$????????????$d_{Tra,Rec}^{t}???t???????????????????$?$???????????$g_{Tra,Rec}^{t}$??????????t????????????????????????x???????
%????????????$t$????????????j?????????????$g_{B,j}^{t}$$,?v?V2I?????????????????$g_{v,B}^{t}$,?v?V2I????d?V2V????????????$g_{v,d}^{t}$???? $v$ ? $d$ ???????? $h_{v,d}^{t} \triangleq 0$,?d?V2V????????????????$g_{d,B}^{t}$??$d^{'}$?V2V?????????????$g_{d,d^{'}}^{t}$?
The indirect channel refers to the virtual channel established through the reflection effect of RIS. Taking the channel gain from the transmitter to the RIS as an example:
%???????RIS?????????????????????RIS????????
\begin{equation}
    g_{Tra,R}^{t} = \sqrt{\rho\beta\left(d_{Tra,R}^{t}\right)^{-\eta}}e^{-j2\pi\frac{d_{Tra,R}^{t}}{\lambda}} \xi_{\text{AoA}}^{t},
\end{equation}
where $\xi_{\text{AoA}}^{t}[l]$ represents the arrival array response from the transmitter to the RIS, defined as: 
\begin{equation}
    \xi_{\text{AoA}}^{t}=[1,\cdots,e^{-j2\pi\frac{d}{\lambda}(F-1)sin(\theta_{AoA}^{t})}],
\end{equation}
%??$\alpha_{\text{AoA}}^{t}[l]$???????RIS???????,????$\alpha_{\text{AoA}}^{t}=[1,\cdots,e^{-j2\pi\frac{d_{Tra,R}^{t}}{\lambda}(F-1)sin(\theta_{AoA}^{t})}]$
where $\lambda$  represents the wavelength of the subchannel. $\theta_{AoA}^{t}$ is the angle of arrival at the $t$  slot, which is the incident angle of the signal reaching the receiving end. When the RIS reflects the received signal, it acts as the transmitting end at this moment. Similarly, the transmitter $Tra \in \{v,d,j,B\}$.
%??$\lambda$??????????d???RIS????????$\theta_{AoA}^{t})$??t??????????????????????RIS??????????????RIS?????

\subsection{Communication Model}
%????
For the V2I link, the actual channel gain of each V2I link can be expressed as $g_{v}^{t} =(g_{R,v}^{t})^{H} \Theta_{t} g_{B,R}^{t}+g_{B,v}^{t}$, and the channel gain for each V2V link pair is $g_{d}^{t} =(g_{R,B}^{t})^{H} \Theta_{t} g_{d,R}^{t}+g_{d,B}^{t}$.

%??V2I?????V2I??????????????$g_{v}^{t} =(g_{R,B}^{t})^{\mathrm{H}} \Theta^{t} g_{v,R}^{t}+g_{v,B}^{t}$?$??V2V?????????$g_{d}^{t} =(g_{R,B}^{t})^{\mathrm{H}} \Theta^{t} g_{d,R}^{t}+g_{d,B}^{t}$

Therefore, at the time $t$, the Signal to Interference plus Noise Ratio (SINR) calculation for the $v$-th V2I link and the $d$-th pair of V2V links is expressed as:
\begin{align}
SINR_{v}^{t} &= \frac{P_{v} \left| g_{v}^{t} \right|^{2}}{\sum\limits_{d \in D} c_{v,d}^{t} P_{d}^{t} \left| g_{d}^{t} \right|^{2} + \sigma_{1}^{2}}, \\
SINR_{d}^{t} &= \frac{P_{d}^{t} \left| g_{d}^{t} \right|^{2}}{\sum\limits_{d \in D} c_{v,d}^{t} P_{d}^{t} \left| g_{d}^{t} \right|^{2} + P_{v} \left| g_{v}^{t} \right|^{2} + \sigma_{1}^{2}},
\end{align}
where $P_{v}$ represents the V2I transmission power, and $P_{d}^{t}$ represents the V2V transmission power in the $t$-th time slot, $\sigma_{1}^{2}$ represents the noise power of communication.
%????t???v?V2I????d?V2V???SINR??????$\gamma_{v}^{t} &= \frac{P_{c} \left| g_{v}^{t} \right|^{2}}{\sum\limits_{d \in D} x_{c,d}^{t} P_{d}^{t} \left| g_{d}^{t} \right|^{2} + \sigma_{a}^{2}}$
%%$\gamma_{d}^{t} &= \frac{P_{d}^{t} \left| g_{v}^{t} \right|^{2}}{\sum\limits_{d \in D} x_{c,d}^{t} P_{d}^{t} \left| g_{d}^{t} \right|^{2} + P_{c} \left| g_{v}^{t} \right|^{2} + \sigma^{2}}.
%??$P_{c}$?V2I?????$P_{d}^{t}$???t???V2V?????$\sigma_{a}^{2}$??????????

Then, the transmission rates for the V2I link and the V2V link are denoted as:
%?????V2I???V2V????????????
\begin{gather}
R_{v}^{t} = W \log_{2}(1 + SINR_{v}^{t}), \\
R_{d}^{t} = W \log_{2}(1 + SINR_{d}^{t}),
\end{gather}
where $W$ is the bandwidth of the spectral subband.
%W????????

\subsection{Sense Model}

%This paper focuses on utilizing communication signals in ISAC cellular systems for target sensing. While transmitting different messages to corresponding vehicular users, the base station (BS) receives and processes the reflected signals from the targets, then uploads the processed signals to the cloud for collaborative analysis and target identification. Let $g_{B,R}^t$, $g_{R,j}^t$, and $g_{B,j}^t$ represent the channel gains from the BS to the RIS, from the RIS to the target, and from the target to the BS, respectively. We define the target vehicle indicator as:

%As shown in Fig.~\ref{fig:enter-label}, the sensing signal is reflected back to the BS by the target via both the direct path and the reflected path. Typically, the propagation delay between these paths is negligible, so the echo channel gain from the target to the BS is expressed as:
As illustrated in Fig.~\ref{fig:enter-label}, the sensing signal is reflected back to the BS by the target vehicle through both a direct path and a reflected path via the RIS. Since the propagation delay difference between these two paths is typically negligible, the echo channel gain from the target to the BS at time slot $t$ can be modeled as \begin{equation}
    g_{j}^{t} =((g_{B,R}^{t})^{H} \Theta_{t} g_{R,j}^{t}+g_{B,j}^{t})((g_{B,R}^{t})^{H} \Theta_{t} g_{R,j}^{t}+g_{B,j}^{t})^{H}.
 \end{equation}
 Accordingly, the sensing SNR for the $j$-th target vehicle is:
%??1?????????????????????????????????????????????[?],?????????????????$H_{j}^{t} =((h_{B,R}^{t})^{\mathrm{H}} \Theta^{t} h_{R,j}^{t}]+h_{B,j}^{t})((h_{B,R}^{t})^{\mathrm{H}} \Theta^{t} h_{R,j}^{t}+h_{B,j}^{t})^{\mathrm{H}}$,???????????????
\begin{equation}
    SNR_{j}^{t} = \frac{P_{j} \left| (g_{j}^{t}) \right|^{2}}{ \sigma_{2}^{2}},
\end{equation}
where $P_{j}$ is the transmission power allocated to the sensing link, and $\sigma_{2}^{2}$ denotes the noise power in the sensing process.
%where,$P_{s}$??????????$\sigma_{b}^{2}$?????????

To guarantee sensing accuracy for each target, the sensing SNR must meet the threshold constraint:
\begin{equation}
    SNR_{j}^{t} \geq SNR_{\text{th}}, \quad \forall j \in J,
\end{equation}
where $SNR_{\text{th}}$ denotes the predefined threshold of SNR.

\subsection{Connectivity and Reliability Model}

To capture the diverse reliability requirements in RIS-assisted ISAC-enabled vehicular networks, we introduce a unified metric termed \emph{CCR}, which accounts for both the temporal continuity of V2I links and the probabilistic reliability of V2V links. We use the term Continuous Connectivity in a generalized sense: for V2I links, it refers to temporal continuity of physical link quality; for V2V links, it refers to the sustained probabilistic availability required for time-critical packet delivery. We define the following metrics:

\subsubsection{Continuous V2I Connectivity (Temporal Reliability)}
To characterize sustained connectivity, we define a continuous connectivity indicator over a sliding window of $N$ consecutive time slots. For vehicle $v$ at time $t$, we define:
\begin{align}
\Psi_{v}^{t} &=
\begin{cases}
1, & \text{if } \sum_{n=t-N+1}^{t} \mathbb{I}\left\{R_{v}^{n} \geq R_{\text{th}}\right\} = N \ \\
0, & \text{otherwise}
\end{cases} \ \\
\Psi_{j}^{t} &=
\begin{cases}
1, & \text{if } \sum_{n=t-N+1}^{t} \mathbb{I}\left\{R_{j}^{n} \geq  
R_{\text{th}}, \text{SNR}_{j}^{n} \geq \text{SNR}_{\text{th}}\right\} = N \ \\
0, & \text{otherwise}
\end{cases}
\vspace{-0.1in}
\end{align}
%For non-sensing vehicles, this requires the downlink transmission rate $R_{v}^{t}$ to exceed a minimum threshold $R_{\text{th}}$.
where $R_{v}^{t}$ and $R_{\text{th}}$ are the transmission rate and the predefine thershold of V2I transmission rate for non-sensing vehicles, respectively. $\Psi_{j}^{t}$ and $\Psi_{v}^{t}$ indicate sustained V2I connectivity for sensing and non-sensing vehicles, respectively. $\mathbb{I}\{\cdot\}$ denotes the indicator function, which returns 1 if the enclosed condition is true and 0 otherwise.%returning 1 if the condition inside is true and 0 otherwise.

%\textbf{1) Instantaneous V2I Connectivity:}  
%For each time slot $t$, a vehicle is considered instantaneously connected to the infrastructure if it satisfies certain quality-of-service thresholds. For non-sensing vehicles, this requires the downlink transmission rate $R_{v}^{t}$ to exceed a minimum threshold $R_{\text{th}}$. For sensing-capable vehicles, both communication and sensing quality must be guaranteed. The instantaneous connectivity indicator is defined as:
%\begin{align}
%\Phi_{v}^{t} &= 
%\begin{cases}
%1, & \text{if } R_{v}^{t} \geq R_{\text{th}}  \\
%0, & \text{otherwise}
%\end{cases}, \\
%\Phi_{j}^{t} &= 
%\begin{cases}
%1, & \text{if } R_{v}^{t} \geq R_{\text{th}} \text{ and } \text{SNR}_{j}^{t} \geq \text{SNR}_{\text{th}} \\
%0, & \text{otherwise}
%\end{cases},
%\end{align}
%where $\Phi_{v}^{t}$ and $\Phi_{j}^{t}$ respectively denote the instantaneous connectivity status of non-sensing and sensing vehicles.

%\textbf{2) Continuous V2I Connectivity (Temporal Domain):}  
%To reflect the sustained link quality required by V2I services, we define a \emph{continuous connectivity indicator} over a sliding window of $N$ consecutive time slots. A vehicle is said to have reliable V2I connectivity at time $t$ if it has remained instantaneously connected throughout the past $N$ slots:
%\begin{align}
%\Psi_{j}^{t} &=
%\begin{cases} 
%1, & \text{if } \sum_{n=t-N+1}^{t} \Phi_{j}^{n} = N \\ 
%0, & \text{otherwise}
%\end{cases}, \\
%\Psi_{v}^{t} &=
%\begin{cases} 
%1, & \text{if } \sum_{n=t-N+1}^{t} \Phi_{v}^{n} = N \\ 
%0, & \text{otherwise}
%\end{cases},
%\end{align}
%where .

\subsubsection{Continuous V2V Connectivity Reliability (Probabilistic Metric)}  
To model the reliability of time-critical V2V transmissions, we adopt a probabilistic success metric based on the likelihood that the entire payload $K$ can be transmitted within the latency constraint $T$. Following~\cite{8792382}, the success probability is expressed as:
\begin{equation}
\operatorname{Pr}\left\{\sum_{t=1}^{T} c_{v,d}^{t} \cdot R_{d}^{t} \cdot \Delta t \geq K \right\},
\end{equation}
where $c_{v,d}^{t} \in \{0,1\}$ denotes whether vehicle $v$ selects channel $d$ at time $t$, $R_{d}^{t}$ is the corresponding achievable transmission rate, and $\Delta t$ is the time slot duration.

\vspace{1mm}
\begin{definition}[Continuous Connectivity Ratio (CCR)]
\emph{CCR} is defined as a unified reliability metric that characterizes both:
\begin{itemize}
    \item \textbf{Temporal reliability of V2I links:} the proportion of vehicles that maintain continuous V2I connectivity over $N$ consecutive time slots while satisfying both communication and sensing thresholds;
    \item \textbf{Probabilistic reliability of V2V links:} the success probability of completing a payload transmission within a strict latency deadline.
\end{itemize}
CCR jointly captures time-extended and task-specific connectivity requirements across heterogeneous V2X services.
\end{definition}

\subsection{Problem Formulation}

To enhance both the reliability of V2V links and the continuous connectivity of V2I links under ISAC constraints, we jointly optimize the channel resource allocation $\mathbf{C}$, transmission power control $\mathbf{P}$, and RIS phase-shift matrix $\boldsymbol{\Theta}$ and maximize CCR: the average delivery success probability of time-critical V2V payloads; the proportion of vehicles maintaining V2I CCR over a sliding window of $N$ slots. The optimization problem is formulated as:
\begin{equation}
\label{eq:problem}
%\mathcal{P}_1:
%\quad
\max_{\{\mathbf{C}, \mathbf{P}, \boldsymbol{\Theta}\}} \frac{1}{T} \sum_{t=1}^{T} \left\{ 
\frac{1}{D} \sum_{d=1}^{D} \operatorname{Pr}\left(  c_{v,d}^{t} \cdot R_{d}^{t} \cdot \Delta t \geq K \right) 
+ \frac{1}{V} \sum_{v \in V} \Psi_v^t 
+ \frac{1}{J} \sum_{j \in J} \Psi_j^t
\right\}
\end{equation}
\begin{align}
\mathbf{s.t.} \quad 
     & R_{v}^{t} \geq R_{\text{th}}, && \forall v \in V,
    \label{eq:comm_constraint} \\
    & SNR_j^t \geq SNR_{\text{th}}, && \forall j \in J,
    \label{eq:sen_constraint} \\
    & c_{v,d}^t \in \{0,1\}, && \forall v \in V, \forall d \in D,
    \label{eq:channel_constraint0} \\
    & \sum_{v \in V} c_{v,d}^t \le 1, && \forall d \in D,
    \label{eq:channel_constraint1} \\
    & P_{d}^{t} \in [P_{\min}, P_{\max}], && \forall d \in D,
    \label{eq:power_constraint} \\
    & \theta_f^i \in \left\{ 0, \tfrac{2\pi}{Q}, \cdots, \tfrac{2\pi(Q-1)}{Q} \right\}, && \forall f \in F.
    \label{eq:phase_constraint}
\end{align}

%\subsection{Problem Formulation}
%To improve the reliability of V2V links and connectivity of V2I links throughout the system, we optimize channel resource allocation $\mathbf{C}$, power control $\mathbf{P}$, and the RIS phase-shift matrix $\boldsymbol{\Theta}$ while ensuring the sensing accuracy of target users, aiming to maximize both the transmission success rate of V2V payloads and the data rate of V2I links. We thus formulate the problem as follows:
%?????????V2V???????V2I????connectivity?????????????????????channel resource allocation, power control, and RIS phase-shift matrix????V2V???????????V2I?????????????????
%\begin{equation} 
%\label{eq:problem}
%\mathbf{P}:
%\max_{\{\mathbf{C}, \mathbf{P}, \boldsymbol{\Theta}\}} \frac{1}{T}\sum_{t=1}^{T}  \left\{ \frac{1}{D}\sum_{d=1}^{D} \operatorname{Pr}\left\{c_{v,d}^{t} \cdot R_{d}^{t} \cdot \Delta t \geq K\right\} + \frac{1}{V}\sum_{v=1}^{V}R_{v}^{t} \right\}
%\end{equation}
%\begin{align}
%    \mathbf{s.t.} \quad 
%    &R_{v}^{t} \geq R_{\text{th}},\forall v \in V,
%    \label{eq:comm_constraint} \\
%    &SNR_j^t \geq SNR_{\text{th}}^{\text{sen}},  \forall j \in J, \label{eq:sen_constraint} \\
%    & x_{v,d}^t \in \{0,1\} \ \forall v \in V,\forall d \in D,
%    \label{eq:channel_constraint0}\\
%    &\sum_{v}^{V} x_{v,d}^t \le  1,  \forall v \in V,
%    \label{eq:channel_constraint1} \\
%    &P_{d}^{t} \in [P_{\min}, P_{\max}], \forall d \in D, \label{eq:power_constraint} \\
%    &\theta_f^i \in \left\{0, \tfrac{2\pi}{Q},\cdots, \tfrac{2\pi(Q-1)}{Q}\right\},\forall f \in F. \label{eq:phase_constraint}
%\end{align}

\section{Proposed VariSAC Framework}
\begin{figure*}[ht!]
\vspace{-0.2in}
    \centering
\includegraphics[width=0.97\linewidth]{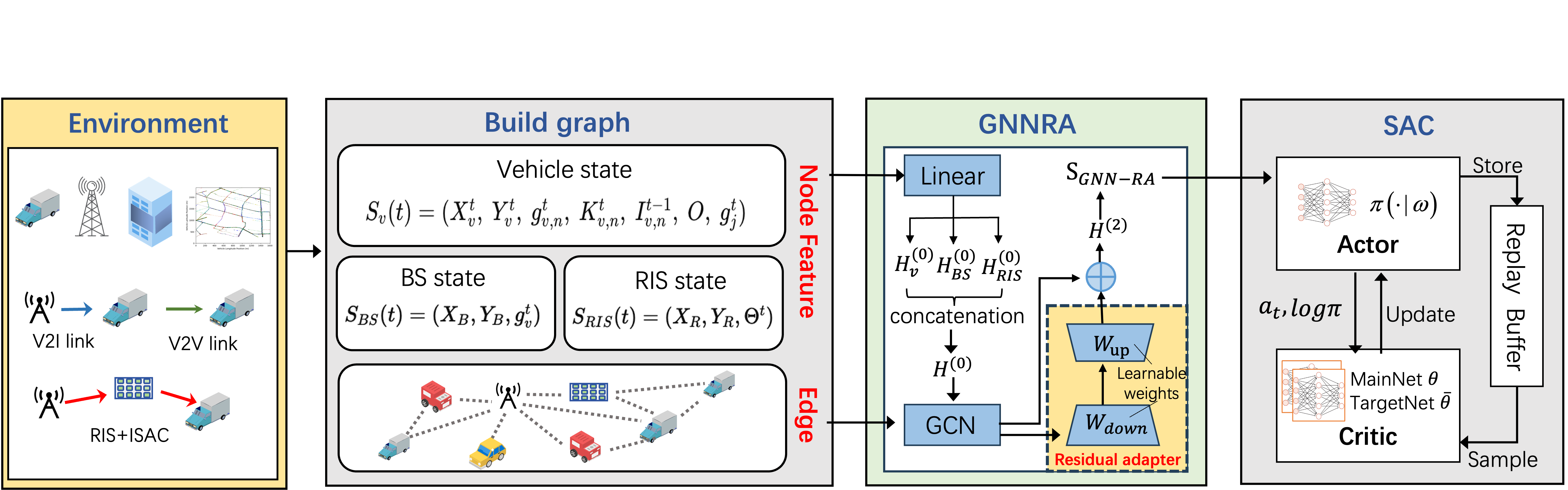}
\vspace{-0.05in}
    \caption{VariSAC. The environment provides heterogeneous states, which are encoded by the GNNRA module. The SAC agent takes the encoded state as input, outputs resource allocation actions, and updates policies through interaction and experience replay.}
    \label{algorithm}
    \vspace{-0.2in}
\end{figure*}

In this section, we introduce the VariSAC framework, which integrates the GNNRA resource encoder and SAC algorithm, as shown in Fig.~\ref{algorithm}. GNNRA module extracts key structural features from the vehicular network, while SAC learns optimal resource allocation policies based on these representations.
\subsection{Markov Decision Process Formulation}
We formulate the resource allocation problem as a Markov Decision Process (MDP), which provides a formal framework widely employed in reinforcement learning to model decision-making under uncertainty.

%We model the resource allocation problem as a Markov Decision Process (MDP), a framework utilized in reinforcement learning techniques.
%????????????????????(MDP)?????????????
\begin{itemize}
    \item  
$\mathbf{State}$: The observed states include three types: vehicle, BS, and RIS states. The observed state of a vehicle includes its coordinates at time slot $t$, its local V2V channel gain $g^t_d$, the remaining load between vehicle v and vehicle n $K_{v,n}^t$, and the interference $I_{v, n}^{t-1}$ at the V2V link receiver at time $t-1$. If the vehicle is also a sensing vehicle, it can observe the sensing channel $g_{j}^{t}$ with $O=1$; otherwise, $g_{j}^{t}=0$ with $O=0$, where $O$ determines whether the vehicle is the target sensing vehicle. The state of the BS involves observing all V2I channels, while the state of the RIS consists of the phase shift of each element. Thus, the complete observed state is as follows: 
% \begin{equation}
%     S_{t}=(S_{1}(t),S_{2}(t),S_{v}(t),\ldots,S_{V}(t),S_{BS}(t),S_{RIS}(t)),
% \end{equation}
%$State$:???????????????????????RIS???????????????t????????????V2V????g???v???n???????$K_{v,n}^t$???t-1??V2V????????I_{\{v, n\}}(t-1)??????????????????????$g_{j}^{t}$??$O=1$,????0???$O$????????????,?????????????V2I???RIS?????????????????????????$s_{t}=(s_{v}(t),s_{B}(t),s_{R}(t))$??$d \in \mathcal{D}$
%$s_{v}(t)=\left(x(t), y(t), g_{\{v, 1\}}(t), g_{\{v, 2\}}(t), g_{\{v, n\}}(t), \ldots, g_{\{v, C\}}(t), I_{\{v, 2\}}(t-1), I_{\{v, n\}}(t-1), \ldots, I_{\{v, C\}}(t-1)\right), \text { if }(v, n) \notin V 2 V \text { link, }g_{\{v, n\}}=0$
\begin{align}
S_{t}&=\Big(S_{v}(t),S_{BS}(t),S_{RIS}(t)\Big),\forall v \in \{1,2,\cdots,V\},\\
S_{v}(t)&= (X_{v}^{t},Y_{v}^{t},g_{v, n}^{t},K_{v, n}^{t},I_{v, n}^{t-1},O,g_{j}^{t}), \forall n \in \{1,2,\cdots,V\},\\
S_{BS}(t)&=(X_B,Y_B,g_{v}^{t}),\forall v \in \{1,2,\cdots,V\},\\
S_{RIS}(t)&=(X_R,Y_R,\Theta^{t}).
\end{align}
% \begin{equation}
% S_{v}(t) = \begin{pmatrix}
% x_{v}^{t},y_{v}^{t},g_{v, 1}^{t},g_{v, 2}^{t},\cdots,g_{v, V}^{t},\\
% K_{v, 1}^{t},\ K_{v, 2}^{t},\cdots,K_{v, V}^{t},\\
% I_{v, 1}^{t-1},I_{v, 2}^{t-1},\cdots,I_{v,V}^{t-1},\ O,\ g_{j}^{t})
% \end{pmatrix}
% \end{equation}

For $S_v(t)$, when $(v,n,t) \notin$ V2V link, $g_{v,n}^t=0$ and $K_{v,n}^t=0$ when $(v, n ,t) \in$ V2V link, $K_{v,n}^t=K_{d}^t$. When $(v,n,t-1) \notin$ V2V link, $I_{v,n}^{t-1}=0$. When $v\notin J$, i.e, vehicle $v$ does not belong to the target sensing vehicle, $g_j^t=0$.

\item $\mathbf{Action}$: BS acts as an agent determining the vehicle's channel allocation, power control, and intelligent reflecting surface phase shift matrix. $c_{v,d}^{t}$ and $p_{d}^{t}$ represent the channel and power selection for the V2V vehicle user pair, respectively, and $\Theta^t$ represents the phase shift selection of the intelligent reflecting surface. Therefore, the action space for time slot $t$ is defined as: 
\begin{equation}
    a_{t}=(c_{v,d}^{t},\ p_{d}^{t},\ \Theta^t).
\end{equation}
%$Action$:??????????????????????????????????????????$x_{v,d}^{t}$?$p_{v,d}^{t}$????V2V??????????????$\Theta^t$???????????????????$t$????????:$a_{t}(x_{v,d}^{t},p_{v,d}^{t},$\Theta^t$)$

\item $\mathbf{Reward}$: In DRL, the design of rewards is flexible, and a well-crafted reward mechanism can enhance system performance. In the resource optimization problem discussed in the previous section, our objectives include two main components: the connectivity of vehicles within the system and the reliability of V2V in the system. Therefore, the reward at time slot $t$ is set as:
%$Reward$:?DRL???????????????????????????????????????????????????????????????????????????V2V??????????????t???????
\begin{equation}
    r_{t}= \sum_{v=1}^{V} \Psi_{v}^{t} + \sum_{j=1}^{J} \Psi_{j}^{t} - \frac{1}{D} \sum_{d=1}^{D} \left( \frac{K_{d}^{t}}{K} \right).
\end{equation}

Through reward, the system maximizes the continuous connectivity of vehicles and penalizes links that fail to complete load transmission, thereby ensuring the connection reliability of V2I and V2V links.

\end{itemize}
%????????????????????????????????????V2I???V2V????????
\subsection{GNNRA Module}
\vspace{-0.03in}
To effectively capture complex structural relationships in the state space, we employ a Graph Convolutional Network (GCN) architecture~\cite{jiang2019semi} to extract essential decision-making information from massive, intricate state data for DRL, while incorporating residual adapters to mitigate gradient vanishing in deep networks, thereby enabling the extraction of multi-level hidden features from high-dimensional state representations. All states and features in this subsection correspond to time step $t$, with temporal notation omitted for readability. 
%???????????????????????????????GCN?????????????????????????DRL???????????????????????????????????????????????????????????????????t??????????????????????????????
The IoV system is modeled as a graph where the BS, RIS, and vehicles serve as nodes, while V2V links, V2I links, and sensing links constitute edges. Each vehicle node is characterized by its complete state vector $S_{veh}$, with BSs and RISs represented by their respective state vectors $S_{BS}$ and $S_{RIS}$.
%?????A????IoV?????????????BS?RIS?????V2V???V2I?????????????????????Sv(t)???????????Sbs(t)??BS??????Sris(t)??RIS??????
Given the inherent dimensionality mismatch among heterogeneous nodes, we project all node features into a unified vector space through the following transformation:
\vspace{-0.05in}
%??????????????????????????????????????????????????
\begin{equation}
\vspace{-0.05in}
    \mathbf{H}_i^{(0)}=Linear_{i}(S_i), \forall i \in \{veh,BS,RIS\},
   % \vspace{-0.03in}
\end{equation}
where $\mathbf{H}_{veh}^{(0)} \in \mathbb{R}^{V \times d}$, $\mathbf{H}_{BS}^{(0)} \in \mathbb{R}^{b \times d}$, $\mathbf{H}_{RIS}^{(0)} \in \mathbb{R}^{r \times d}$ and $Linear_i$ is a single-layer linear layer built for different types of nodes $i$. 

Then, we employ standard GCN layers to aggregate information across nodes. GCN has demonstrated superior information extraction capabilities and has been widely adopted in communication systems. Node features propagate and iteratively update through link-based message passing as follows: 
%??????????GCN??????????????GCN????????????????????????????IoV???????????????????????????????????
\begin{align}
\vspace{-0.05in}
\mathbf{H}^{(0)}&=Concat(\mathbf{H}_{veh},\mathbf{H}_{RIS},\mathbf{H}_{BS}),\\
    \mathbf{H}^{(1)}&=\sigma(\mathbf{\tilde A}\mathbf{H}^{(0)}\mathbf{W}^{(0)}),
    \vspace{-0.05in}
\end{align}
where $Concat(\cdot)$ denotes vector concatenation operation and $\mathbf{H}^{(0)} \in \mathbb{R}^{(V+b+r) \times d}$. $\mathbf{\tilde A}$ is the normalized adjacency matrix and $\mathbf{W}^{(0)} \in \mathbb{R}^{d \times d_b}$ is a learning weight matrix. $\sigma$ is the activation function, uniformly using ReLU in the experiment. To enhance representation capacity and mitigate oversmoothing, we insert a residual adapter module between GCN layers:
\vspace{-0.03in}
%???????????????????????????GCN?????????????????
\begin{equation}
f_{\text{RA}}(\mathbf{H}^{(1)}) =\sigma(\mathbf{H}^{(1)}\mathbf{W}_{\text{down}})\mathbf{W}_{\text{up}}, 
\vspace{-0.03in}
\end{equation}
where $\mathbf{W}_{\text{down}} \in \mathbb{R}^{d_b \times d_h}$, $\mathbf{W}_{\text{up}} \in \mathbb{R}^{d_h \times d_b}$ are learnable weight matrices and $d_b \ll d_h$ is the bottleneck dimension. The residual-enhanced feature is fed into the second GCN layer:
%?? $\mathbf{W}_{\text{down}} \in \mathbb{R}^{d_h \times d_b}$?????, $\mathbf{W}_{\text{up}} \in \mathbb{R}^{d_b \times d_h}$ ??????? $d_b \ll d_h$??????????????????????GCN??
\begin{equation}
\mathbf{H}^{(2)} = \mathbf{\tilde{A}} (\mathbf{H}^{(1)} + f_{\text{RA}}(\mathbf{H}^{(1)})) \mathbf{W}^{(1)},
\end{equation}
where $\mathbf{H}^{(2)} \in \mathbb{R}^{(V+b+r) \times d'}$ and $\mathbf{W}^{(1)} \in \mathbb{R}^{d_b \times d'}$ is learnable matrix. Finally, the output features of all nodes are flattened to produce the state representation used by the DRL agent:
%?????????????????????????DRL???????????????
\begin{equation}
S_{GNNRA} = Flatten(\mathbf{H}^{(2)}),
\end{equation}
where $S_{GNNRA} \in \mathbb{R}^{1 \times (V+b+r)d'}$ and $Flatten(\cdot)$ is the operation of converting a vector into a 1-dimensional vector.

\subsection{SAC Training}
The state representation encoded by the GNNRA module serves as the input to the SAC algorithm. SAC leverages a maximum entropy objective to improve policy stability and sample efficiency, while its dual Q-learning structure and soft value function further enhance optimization reliability. Thanks to its robustness to hyperparameter choices, SAC offers stable and flexible performance across diverse, high-dimensional control tasks, making it well-suited for dynamic vehicular network scenarios.

At each training step $t$, the objective is to maximize both the expected cumulative reward and policy entropy:
\begin{align}
\hspace{-0.15in}J(\pi(a|s)) =
\mathbb{E}_{\tau \sim \pi}\left[
\sum_{t=0}^{T} \gamma^{t} r(s_t, a_t)
+ \alpha H(\pi_\theta(\cdot|s_t))
\right],
\end{align}
where $\gamma$ is the discount factor and $\alpha$ is a temperature parameter balancing reward and exploration.

Experience tuples $(s_n, a_n, r_n, s_{n+1})$ are stored in a replay buffer and sampled in mini-batches to update the networks. The temperature parameter $\alpha$ is updated via:
\begin{equation}
\nabla_{\alpha}J(\alpha) = \mathbb{E}_{(s,a)\sim \pi}\left[ -\log \pi\varphi(a|s) \right].
\end{equation}

The critic networks are updated using the target:
\begin{align}
Q_{\text{tar}}(s_{n+1}, a_{n+1}) = \min_{i=1,2} Q_{\bar{\theta}i}(s{n+1}, a_{n+1}) - \alpha \log\pi_\theta(a_{n+1}|s_{n+1}),
\end{align}
with the loss defined as the mean squared error between $Q_{\theta_i}(s_n, a_n)$ and $r_n + \gamma Q_{\text{tar}}(s_{n+1}, a_{n+1})$.

The actor is updated via the policy gradient, where actions are sampled as $a_{n, \text{new}} = f_\phi(\zeta_n; s_n)$ with $\zeta_n \sim \mathcal{N}(0, I)$:
\begin{align}
\nabla_\phi J(\phi) = \nabla_\phi \alpha \log\pi_\phi(a_{n,\text{new}} | s_n) - \nabla_\phi Q(s_n, a_{n,\text{new}}).
\end{align}
 The gradients of the Adam optimizer based on the aforementioned loss function are used to update the parameters $\alpha$, $\theta_1$, $\theta_2$, and $\phi$. Finally, the parameters $\bar{\theta}_1$ and $\bar{\theta}_2$ of the two target critic networks will be updated to:
%????????????Adam???????????????????????????????????????????????
\begin{align}
\bar{\theta}_1 &= \tau_1 \theta_1 + (1 - \tau_1) \bar{\theta}_1, \\
\bar{\theta}_2 &= \tau_2 \theta_2 + (1 - \tau_2) \bar{\theta}_2,
\end{align}
where $\tau_1$ and $\tau_2$ are constants satisfying $\tau_1 \ll 1$ and $\tau_2 \ll 1$. After all parameters are updated, the algorithm will proceed to process the next batch of data, and the optimal policy will be obtained upon completion of training.

\section{Performance Evaluation}

\begin{figure*}
    \centering
    \vspace{-0.1in}
    \includegraphics[width=1\linewidth]{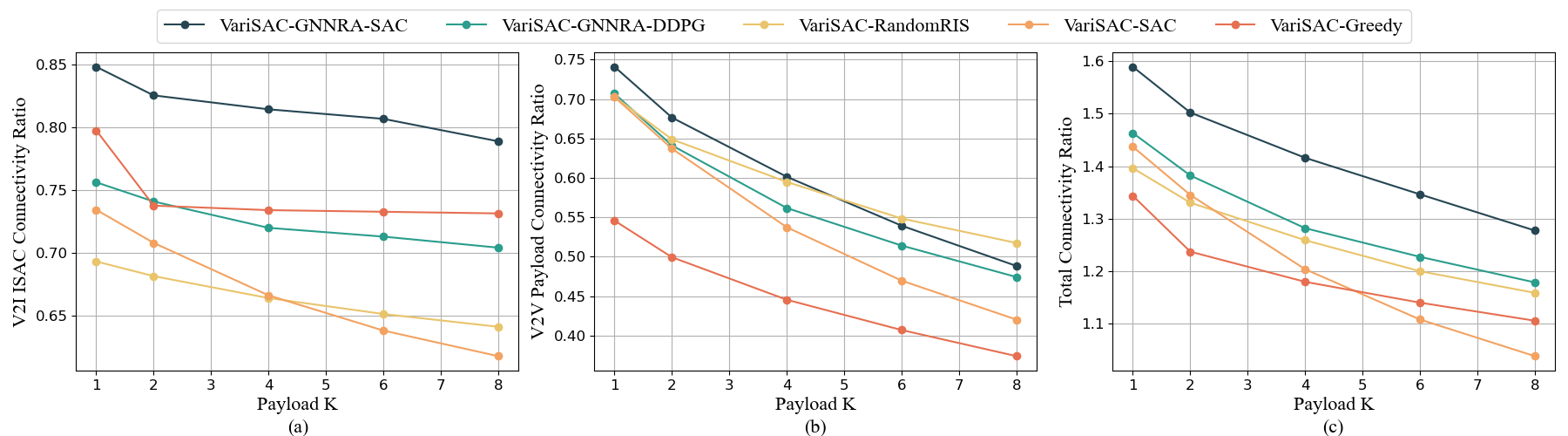}
    \vspace{-0.15in}\caption{Comparison of the effects of different payload $K$ on the three connectivity ratios.}
    \label{fig:payload_K}
\end{figure*}

\begin{figure*}
    \centering
    \vspace{-0.1in}
    \includegraphics[width=0.95\linewidth]{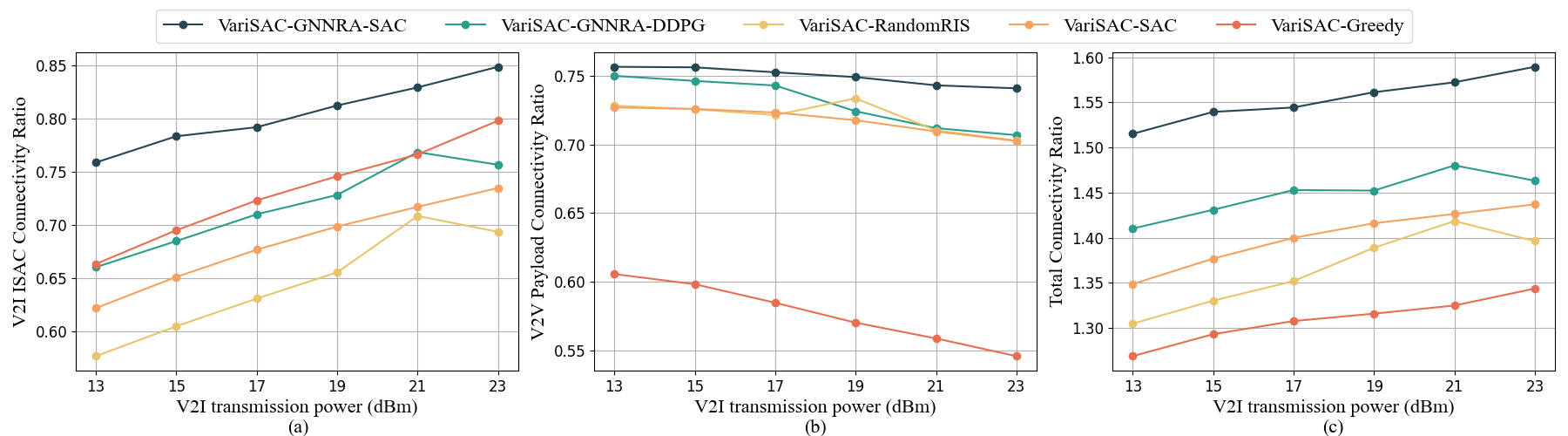}
    \vspace{-0.15in}\caption{Comparison of the effects of different V2I transmission power on the three connectivity ratios.}
    \label{fig:V2I_power}
\end{figure*}

\subsection{Parameters Setup}
%The path loss model for simulated data transmission is based on references~\cite{harounabadi2021v2x,ii2007winner}. This paper considers a scenario comprising one base station, one RIS, and twelve vehicles, two of which are target vehicles. Each vehicle establishes full connections with both the base station and the RIS, and maintains a single connection with its neighboring vehicle.The base station is located at the three-dimensional coordinates $(180, 270, 25)$ meters, and the RIS is positioned at $(290, 380, 25)$ meters. The vehicles are uniformly distributed on a road network spanning $650$ meters in length and $450$ meters in width, covering four directions (up, down, left, and right), with each direction featuring three dual-lane roads. The initial vehicle speeds are randomly selected from the range [10,15] m/s, and the initial V2V load is 8×1060 bits. The RIS reflecting surface consists of $F=12$ reflective elements with a phase shift quantization of $Q=8$. The V2I transmission power is $P_v = 23$ \text{dBm}, the minimum transmission rate is \( 3\ \text{bps/Hz} \), the sensing link transmission power is $P_j = 23$ \text{dBm}, the sensing SNR threshold is 10 \text{dB}, the V2V link transmission power is $P_d = [1,\ 2,\ \ldots,\ 23]$ \text{dBm}, the noise powers are $\sigma_1, \sigma_2 = -114$ \text{dBm}, and each subchannel in the communication is allocated a bandwidth of 1 MHz.
The path loss model follows~\cite{harounabadi2021v2x,ii2007winner}. The scenario includes one base station, one RIS, and twelve vehicles (two as targets). Each vehicle is fully connected to both the base station and the RIS, and has a single V2V link to its neighbor. The base station and RIS are located at $(180, 270, 25)$ m and $(290, 380, 25)$ m, respectively. Vehicles are uniformly distributed over a $650 \times 450$ m$^2$ road network with four directions (up, down, left, right), each comprising three dual-lane roads. Initial vehicle speeds are randomly chosen from $[10, 15]$ m/s, with a V2V load of $8 \times 1060$ bits. The RIS consists of $F = 12$ elements with $Q = 8$ phase quantization. V2I and sensing transmission powers are both $23$ dBm, the minimum transmission rate is $3$ bps/Hz, and the sensing SNR threshold is $10$ dB. V2V transmission powers are set as $P_d = [1,,2,\ldots,23]$ dBm, with noise powers $\sigma_1, \sigma_2 = -114$ dBm. Each communication subchannel has a $1$ MHz bandwidth.
%(1)System Setup: ??????????????????[1][2]???????????????RIS?????????????????????????(180, 270, 25)???RIS??(290, 380, 25)???????????650???450??????????????????????????????????????[10,15]m/s???????V2V?????8×1060 bit?RIS?????F=12??????????????Q=8?V2I?????$P_v=23dBm$,???????$3 bps/Hz$,?????????$P_j=23dBm$,??SNR???$15dB,$V2V???????$P_d=[1,2,\ldots,23]dBm$,????$\sigma_1,\sigma_2=-114dBm$,??????????????1MHz?
%[1]3rd Generation Partnership Project (3GPP), ?Study on channel model for frequencies from 0.5 to 100 GHz, release 16,v16.1.0,? 3GPP, Sophia Antipolis, France, Tech. Rep. TR 38.901, Dec. 2019.
%[2]?WINNER II channel models,? document IST-4-027756 WINNER II D1.1.2 V1.2, Sep. 2007. [Online]. Available: http://projects.celticinitiative.org/winner/WINNER2-Deliverables/D1.1.2v1.1.pdf

The GNNRA framework employs a multi-layer GNN with two main modules: a node encoding layer?using three independent NodeEncoder networks for vehicles, RIS, and BS?and a two-layer graph convolution module. The first GCN layer is followed by a bottleneck residual module; the second outputs the final node features. The SAC algorithm uses one actor, two critics, and two target critics, each as a five-layer fully connected network (three hidden layers of 512 neurons). Training uses 1,800 episodes, a discount factor $\gamma = 0.99$, and seed 12345. Testing is performed on real vehicle trajectories, with 100 time slots per test and results averaged over 50 runs.
%The GNNRA framework adopts a multi-layer graph neural network architecture, consisting of two main modules: the node encoding layer and the graph convolution layer. The node encoding layer employs three independent NodeEncoder networks to process vehicle, RIS, and BS nodes, respectively. The core part of the graph convolution features a two-layer structure: the first GCN layer is followed by a residual module with a bottleneck structure, while the second GCN layer outputs the final node features.  
\begin{table}[h]
  \centering
  \caption{Neural Network Parameters}
  \label{tab:nn_params}
  \begin{tabular}{ll}
    \toprule
    \textbf{Parameter} & \textbf{Value} \\
    \midrule
    Optimizer & Adam \\
    Discount factor $\gamma$ & 0.99 \\
    Critic/Actor networks learning rate & $3\text{e}{-4}$ \\
    Nonlinear function & ReLU \\
    Experience replay buffer size $\mathcal{B}$ & $10^{6}$ \\
    Mini-batch size $I$ & $10^{6}$ \\
    Number of episodes & 1800 \\
    Number of iterations per episode & 100 \\
    Size of hidden layers & 512 \\
    Target network soft update parameters $\tau_{1}$, $\tau_{2}$ & 0.01 \\
    \bottomrule
  \end{tabular}
\end{table}

%The SAC algorithm's structure includes one actor network, two critic networks, and two target critic networks. All networks utilize a five-layer fully connected neural network architecture, comprising one input layer, one output layer, and three hidden layers, with each hidden layer containing 512 neurons. During training, 1,800 training episodes are set, with the discount factor $\gamma$ set to 0.99, seed is 12345. After training, testing is conducted based on the trained model and real vehicle trajectories, with each test round consisting of 100 time slots. The final results are obtained by averaging the outcomes of 50 sequential test runs.
%(2)Algorithm Setup: GNNRA?????????????????????????????????????????????NodeEncoder?????????RIS?BS????????????????????GCN?????????????????GCN?????????SAC??????????????????????????????????????????????????????????????????????????????????????512?????????1800?????????????0.99??????????????????????????????100????????????50????????????

%(3)Metric Setup: ????????????????????????????????????????????V2V??????????

We proposed several baseline models for comparison:
%?????????????????GNNSAC???????
\begin{itemize}
    \item \textbf{DDPG}: To evaluate the advantages of employing the SAC algorithm in our VariSAC framework, we replace SAC with the DDPG algorithm~\cite{lillicrap2015continuous}. Indicated as VariSAC-GNNRA-SAC.
    \item \textbf{SAC}: This scheme uses only the SAC algorithm for decision-making to evaluate the performance improvement provided by GNNRA. Indicated as VariSAC-SAC.
    %?????????RGNN????????????SAC?????????
    \item \textbf{Random RIS}: The scheme sets except for setting the RIS phase shift matrix randomly, all other choices are consistent with the VariSAC algorithm. Indicated as VariSAC-RandomRIS.
    %??????RIS?????????????????VariSAC ???????
    \item \textbf{Greedy}: This baseline selects the action that maximizes the immediate reward at each time step, without considering future impact or environmental dynamics. Indicated as VariSAC-Greedy.
    %??????????????????????????????????????
\end{itemize}

The main evaluation metrics correspond directly to  CCR:
%\begin{itemize}
%    \item V2I ISAC Connectivity Ratio: The ratio of the number of continuous successful connections to the total number of connections within the T time slot.Calculated by $\frac{\frac{1}{V}\sum_{t=N}^{T}\Psi_{v}^{t}}{T-N+1}$.
%    \item V2V Payload Connectivity Ratio: The success probability of transmitting load $K$ within $T$ time slots.
%    \item Total Connectivity Ratio: The connectivity of the entire system consists of V2I ISAC Connectivity Ratio and V2V Payload Connectivity Ratio.
%\end{itemize}
\begin{itemize}
\item \textbf{V2I ISAC Connectivity Ratio (V2I CCR):} The fraction of time slots in which vehicles maintain uninterrupted connectivity over a sliding window of $N$ slots:
\begin{equation}
\mathrm{CCR}_{\mathrm{V2I}} = \frac{1}{V(T-N+1)} \sum_{v=1}^{V} \sum_{t=N}^{T} \Psi_{v}^{t},
\end{equation}

\item \textbf{V2V Payload Connectivity Ratio (V2V CCR):} The success probability that the payload $K$ can be delivered within $T$ time slots on the V2V link, i.e., the continuous connectivity reliability for V2V transmission:
\begin{equation}
\text{CCR}_{V2V} = \operatorname{Pr}\left\{\sum_{t=1}^{T} c_{v,d}^{t} R_{d}^{t} \Delta t \geq K \right\}.
\end{equation}

\item \textbf{Total Connectivity Ratio (Total CCR):} The total continuous connectivity reliability of the system, calculated as the sum of V2I CCR and V2V CCR:
\begin{equation}
\text{CCR}_{\text{Total}} = \text{CCR}_{V2I} + \text{CCR}_{V2V}.
\end{equation}
%????????????????????
\end{itemize}

%\vspace{-0.1in}
\subsection{Convergence Performance}
\begin{figure}
    \centering
     \vspace{-0.2in}
     \includegraphics[width=0.93\linewidth]{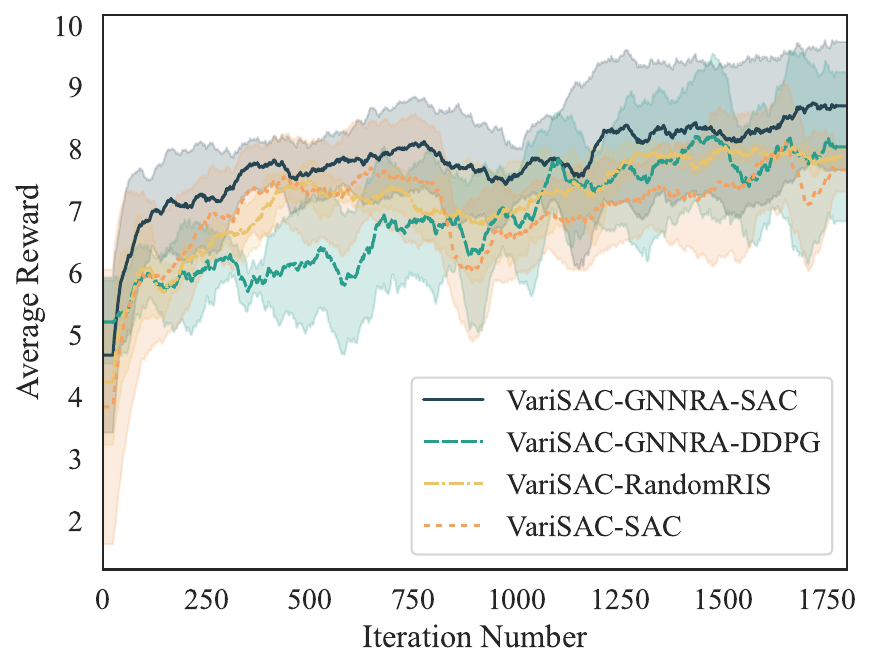}
    \vspace{-0.15in}
    \caption{Compare the impact of VariSAC and different baselines on the training process rewards at different learning ratios.}
    \label{fig:reward}
    \vspace{-0.2in}
\end{figure}

We first compare the training reward curves of our proposed VariSAC method against various baselines under different learning rates $\{0.0001, 0.0003, 0.0005\}$ to evaluate robustness. To clearly illustrate reward progression, smoothing techniques are applied and variance is plotted. %To better illustrate reward evolution, we apply smoothing and plot the variance. 
As shown in Fig.~\ref{fig:reward}, VariSAC consistently outperforms all baselines, exhibiting the fastest convergence and the highest sustained reward throughout training. Notably, the removal of the GNNRA module leads to a pronounced decline in reward, highlighting its critical contribution to overall performance.

\subsection{Simulation Results}
%???????????
\vspace{-0.07in}
\subsubsection{Impact of the Payload $K$} 
We set the payload $K \in [1,2,4,6,8]$ to verify changes in algorithm performance under different scenarios. As shown in Fig.~\ref{fig:payload_K}, our proposed method, VariSAC, achieved the best results in almost three connectivity ratio metrics. For the total connectivity ratio, VariSAC-GNNRA-SAC achieves remarkable improvement over VariSAC-GNNRA-DDPG by $8.38\%$, VariSAC-RandomRIS by $10.24\%$, VariSAC-SAC by $10.61\%$, and VariSAC-Greedy by $15.54\%$. As the payload increases, resource allocation becomes more difficult under resource-constrained conditions, leading to a corresponding decrease in the connectivity ratio. In the baselines, VariSAC-GNNRA-DDPG achieved the suboptimal results, as changing the DRL method to VariSAC has the least impact compared to other baselines. Note that for the V2V payload connectivity ratio, VariSAC-RandomRIS is slightly higher than our proposed method when $K=6,8$. It is because, in the case of random RIS, SAC can only optimize channel and power selection, which enables it to achieve relatively stable connectivity on the V2V link, but completely ignores the connectivity reliability of the V2I and perception links. Fig.~\ref{fig:payload_K}(a) proves this point.

\subsubsection{Impact of the V2I Power} 
We vary the V2I transmission power $P_v \in [13,15,17,19,21,23]$ to evaluate the algorithm performance under different power levels. As shown in Fig.~\ref{fig:V2I_power},  increasing the V2I transmission power enhances the transmission rate, thereby facilitating more reliable V2I link connections. However, this also introduces greater interference to V2V links, resulting in a slight decline in the V2V payload connectivity ratio for VariSAC.
Despite this, the total connectivity ratio of VariSAC continues to rise steadily as the V2I power increases, since the improvement in V2I ISAC connectivity significantly outweighs the reduction in V2V payload connectivity. For the total connectivity ratio, VariSAC-GNNRA-SAC achieves remarkable improvement over VariSAC-GNNRA-DDPG by $6.22\%$, VariSAC-RandomRIS by $10.84\%$, VariSAC-SAC by $10.23\%$, and VariSAC-Greedy by $18.10\%$. In contrast, VariSAC-Greedy seeks to optimize immediate performance at each time slot. Given that the reward function assigns a higher weight to the V2V ISAC connectivity ratio and a lower weight to the V2V payload connectivity ratio, Greedy tends to prioritize the former?often at the cost of V2V link stability.
%with the increase in V2I transmission power, its transmission rate increases, making it easier to achieve reliable connections of the V2I link. However, the increase in V2I transmission power also increases the interference of V2I links on V2V links, causing the V2V payload connectivity ratio of VariSAC to decrease slightly. For the total connectivity ratio, since the increase rate of V2I ISAC connectivity far exceeds the decrease rate of V2V payload connectivity ratio, VariSAC also shows a steady increase trend as V2I transmission power increases. In contrast, VariSAC-Greedy aims to select the optimal solution at each time slot, and the V2V ISAC connectivity ratio has a relatively large weight in the reward, while the V2V payload connectivity ratio has a relatively small weight. Therefore, Greedy focuses more on optimizing the V2V ISAC connectivity ratio and sacrifices the connection stability of V2V links.

\begin{figure}
    \centering
    \vspace{-0.1in}
    \includegraphics[width=0.93\linewidth]{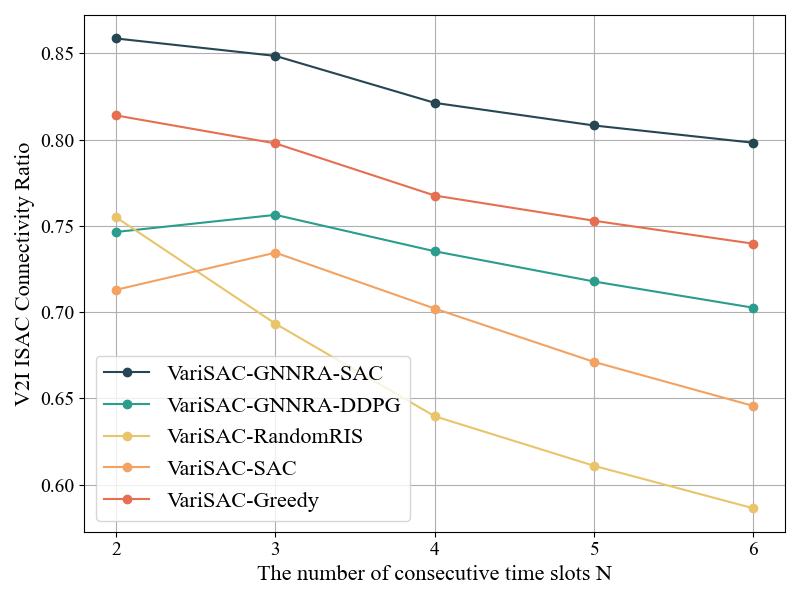}
    \vspace{-0.1in}
    \caption{Comparison of the effects of the number of consecutive time slots $N$ on the continuous connectivity ratios.}
    \label{fig:consecutive_time_slots_N}
\end{figure}

\begin{figure}
    \centering
     \vspace{-0.1in}
     \includegraphics[width=0.95\linewidth]{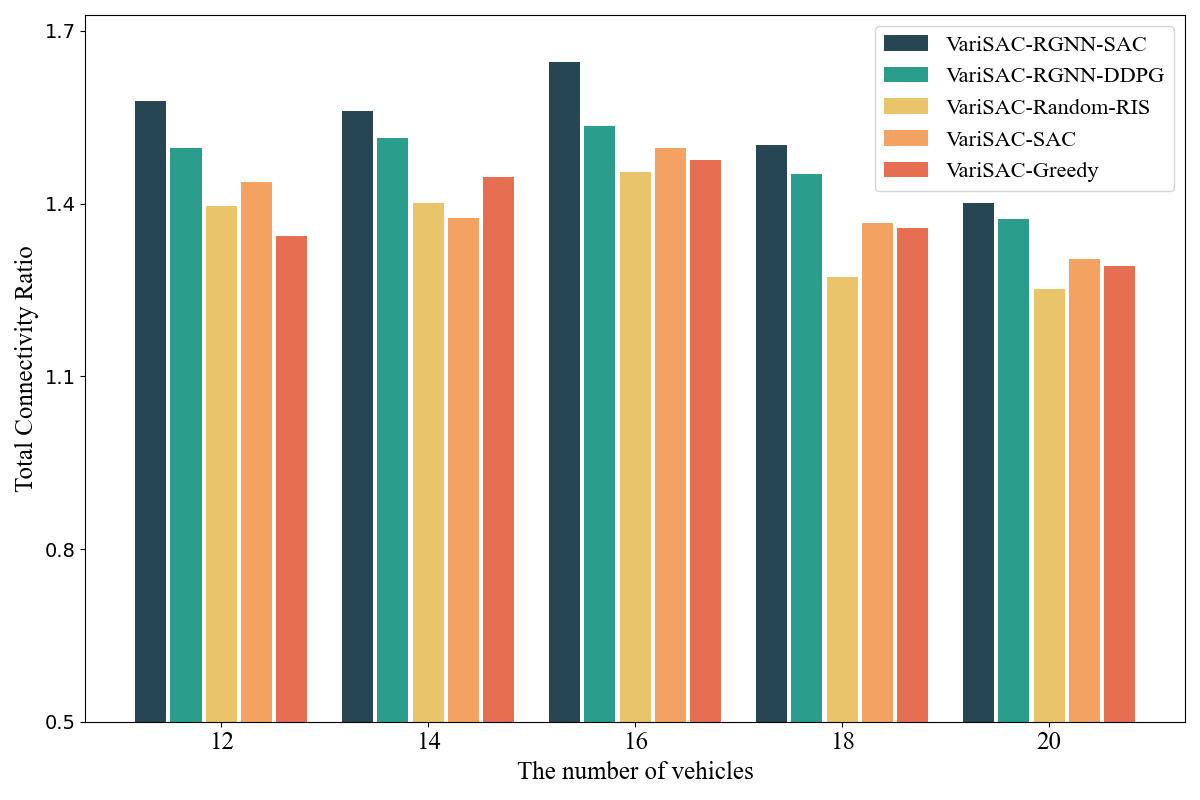}
    \vspace{-0.1in}
    \caption{Comparison of the effects of the number of vehicles on the reliable connectivity ratios.}
    \label{fig:vehicle_numbers}
    \vspace{-0.15in}
\end{figure}

\subsubsection{Impact of the Number of Consecutive Time Slots} 
To conduct a fine-grained evaluation of our method's superiority, we calculated the V2V ISAC connectivity ratio for different values of continuous time slot $N\in [2,3,4,5,6]$. As shown in Fig.~\ref{fig:consecutive_time_slots_N}, the V2I ISAC connectivity ratio of VariSAC gradually decreases as the number of consecutive time slots increases. The reason for the decrease is that maintaining connection reliability in long time slots requires more stringent resource allocation decisions. However, VariSAC still achieves a significant advantage across all slot lengths. For the V2I ISAC connectivity ratio, VariSAC-GNNRA-SAC achieves remarkable improvement over VariSAC-GNNRA-DDPG by $11.70\%$, VariSAC-RandomRIS by $13.74\%$, VariSAC-SAC by $15.52\%$, and VariSAC-Greedy by $5.48\%$. VariSAC-RandomRIS, due to its random nature, allocates resources in a more dispersed manner, resulting in a rapid decline in performance in the case of long continuous slots. The results of other baselines are consistent with Fig.~\ref{fig:payload_K}(a) and Fig.~\ref{fig:V2I_power}(a).

\begin{figure*}
    \centering
   \vspace{-0.1in}
   \includegraphics[width=0.95\linewidth]{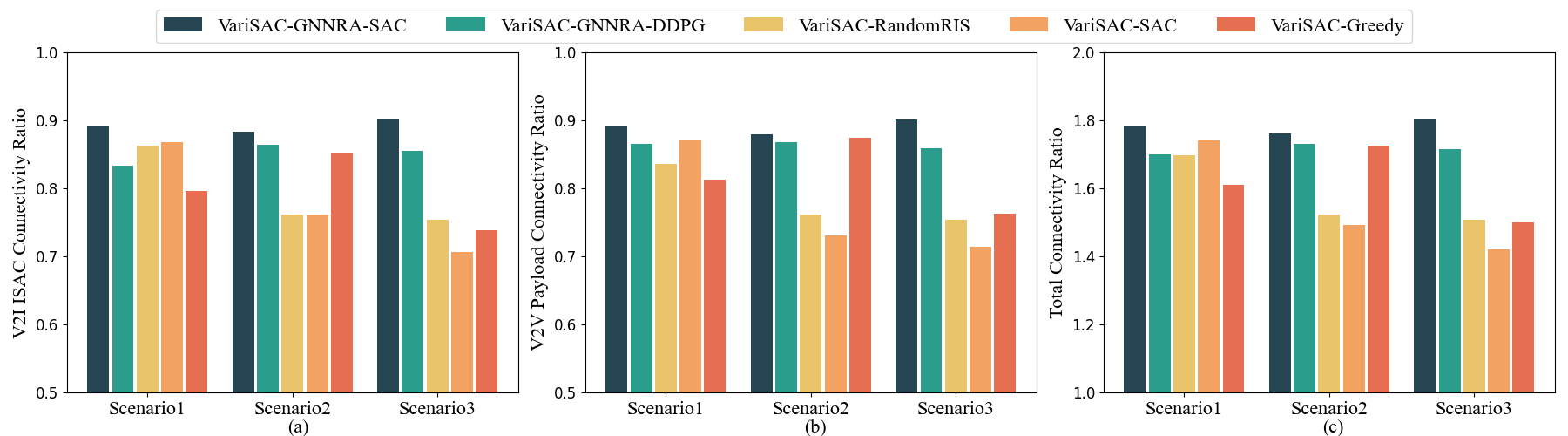}
    \vspace{-0.2in}
    \caption{Comparison of the effects of different scenarios on the three connectivity ratios.}
    \label{fig:dif_scenarios}
    \vspace{-0.18in}
\end{figure*}

\subsubsection{Impact of the Number of Vehicles}
As the number of vehicles increases, the search space for DRL-based resource allocation decisions expands dramatically. We set the number of vehicles to [12, 14, 16, 18, 20] to compare the performance of different models in complex scenarios. As shown in Fig.~\ref{fig:vehicle_numbers}, the performance of VariSAC increases slightly at first and then decreases rapidly as the number of vehicles increases. We analyzed the reason as follows: when the number of vehicles increases from 12 to 16, the distance between vehicles decreases, channel interference reduces, and links in the system become easier to establish reliable connections. However, as the number of vehicles continues to increase, the search space for resource allocation expands rapidly, making it difficult to maintain a high connection reliability rate. Despite this, VariSAC consistently outperforms the other models across all scenarios. For the V2V connectivity ratio, VariSAC-GNNRA-SAC achieves remarkable improvement over VariSAC-GNNRA-DDPG by $2.06\%$, VariSAC-RandomRIS by $11.42\%$, VariSAC-SAC by $7.47\%$, and VariSAC-Greedy by $7.86\%$.

%, demonstrating its robustness and scalability in increasingly complex environments.%VariSAC achieved the best performance in all scenarios.
%VariSAC-sac???????????
% ??????[12,14,16,18,20]???

\subsection{Validation with Real-World Trajectories}
%Thj??????????????????????????????
We evaluated the proposed method using real-world data from the Didi GAIA Open Dataset, specifically focusing on vehicle trajectories recorded between 8:00 and 8:05 AM on November 16, 2016. To simulate realistic vehicle mobility patterns, these trajectories were mapped onto our simulated environment. Multiple trajectory scenarios were then generated using different sampling strategies. The connection ratio was used as the primary performance metric. Three distinct scenarios were constructed: Scenario 1 adopts random sampling, Scenario 2 uses area-balanced sampling, and Scenario 3 prioritizes long trajectories.

%We used real-world data from Didi GAIA Open Data, specifically selecting trajectories from 8:00 to 8:05 AM on November 16, 2016. To simulate realistic vehicle movement patterns, we mapped these real trajectories onto our simulated environment and collected multiple trajectory scenarios using different sampling methods. We evaluated the performance using the connection ratio as the key metric. Scenarios 1, 2, and 3 correspond to random sampling, area-balanced sampling, and long-trajectory priority sampling, respectively.% The scenarios  are random collection, area-balanced collection, and long-trajectory priority selection.
%???????Didi GAIA Open Data???????2016?11?16???8??8?05??????????????????????????????????????????????????????????????????????????????????????????????????????

The aggregated experimental results are presented in Fig.~\ref{fig:dif_scenarios}. For the total connectivity ratio, VariSAC consistently outperforms the other models across all scenarios. For the V2V connectivity ratio, VariSAC-GNNRA-SAC achieves remarkable improvement over VariSAC-GNNRA-DDPG by $1.81\%$, VariSAC-RandomRIS by $5.08\%$, VariSAC-SAC by $2.57\%$, and VariSAC-Greedy by $2.13\%$. These results demonstrate that the introduction of GNNRA leads to significant performance improvements, particularly in Scenarios 2 and 3. This indicates that GNNRA effectively captures the complex spatio-temporal relational structures among vehicles and utilizes them for more optimal resource allocation decisions. Furthermore, the comparison with DDPG highlights that the SAC algorithm is better suited for resource allocation tasks, owing to its capability to handle continuous action spaces and enhance exploration efficiency. The consistently poor performance of random RIS selection across all scenarios underscores the effectiveness of VariSAC in exploiting the auxiliary benefits of RIS. Notably, in long-trajectory scenarios, VariSAC exhibits a more substantial performance advantage over the baselines, suggesting its superior ability to manage complex and extended vehicle trajectories while ensuring continuous communication reliability.
%The experimental results of DDPG show that, compared to DDPG, the SAC algorithm is more suitable for resource allocation problems due to its ability to handle continuous action spaces and improve exploration efficiency. The poor performance of the random RIS across all scenarios also reflects that VariSAC can significantly enhance the auxiliary gain of the RIS. In long-trajectory scenarios, VariSAC demonstrates a more pronounced advantage over the baseline, indicating that VariSAC performs exceptionally well in handling more complex and longer vehicle trajectories while meeting continuous communication requirements.

\section{Related Work}
RIS enhance the coverage, capacity, and energy efficiency of communication systems through intelligent reflection control of channel environments. However, the high-dimensional nonlinear characteristics of their phase optimization problems provide a natural scenario for deep learning applications, where deep learning methods can achieve real-time adaptive beamforming in complex environments through data-driven approaches. Sheen \textit{et al}.~\cite{9317827} proposed a DL-based RIS phase configuration optimization method that significantly improves the spectral efficiency of wireless communication systems by leveraging the interactive features between receiver location attributes and RIS reflection beamforming vectors, achieving near-theoretical upper bound performance with only a small amount of training data. Zahedi \textit{et al}.~\cite{9827185} presented a DL-based RIS beamforming design that optimizes RIS element phase shifts in real-time through neural networks, significantly enhancing spectral efficiency and energy efficiency of 6G communications in low SNR scenarios while overcoming the high computational complexity limitations of traditional optimization methods. Dinh-Van \textit{et al}.~\cite{10570724} proposed an unsupervised DL-based distributed RIS joint beamforming design method that simultaneously optimizes base station active beamforming and multi-RIS passive phase shift configurations through end-to-end neural networks, achieving  higher spectral efficiency than centralized RIS in industrial scenarios while reducing computational latency by two orders of magnitude. Sagir \textit{et al}.~\cite{sagir2023deep} presented a model-driven deep learning-based joint beamforming design for RIS-assisted multicast systems, which unfolds the alternating projection gradient algorithm into a trainable neural network to maximize total rate while ensuring RIS element modulus constraints. Ding \textit{et al}.~\cite{10419704} introduced a DL-based cooperative RIS communication system for IoT that achieves joint optimization of RIS phase optimization and symbol detection through a dual neural network architecture, reducing bit error rate to 1/10 of traditional methods under Rayleigh fading channels while enabling distributed cooperative control of RIS elements through IoT networks. However, most existing optimization approaches are designed for static or quasi-static environments and depend heavily on large volumes of labeled data, resulting in limited generalization. In highly dynamic settings with high-mobility users or rapidly varying channels, these methods struggle to adapt to environmental changes, leading to suboptimal performance and inefficient optimization.

To address this problem, DRL, as a method that requires no large amounts of labeled data and learns autonomously through environmental interaction, provides new insights for dynamic RIS-assisted communication scenarios. Unlike DL, DRL can interact with the environment and continuously optimize decision strategies through trial-and-error mechanisms, effectively handling time-varying channels and mobile user scenarios. In recent years, numerous researchers have explored the application potential of DRL in RIS-assisted communication systems. Hazarika \textit{et al}.~\cite{10017765} proposed a MA-DRL based reconfigurable intelligent surface-assisted vehicular network system that significantly improves three core performance indicators: communication quality, computational efficiency, and network intelligence through joint optimization of RIS phase shifts and computation offloading strategies.  Hu \textit{et al}.~\cite{10901761} presented an MA-DRL-based multi-RIS-assisted millimeter-wave D2D secure communication system that achieves excellent performance in security, communication quality, and resource utilization through intelligent optimization of RIS configuration and resource allocation. Hazarika \textit{et al}.~\cite{10261304} proposed an MA-DRL-based RIS-assisted vehicular network system that achieves communication quality improvement, resource efficiency optimization, and task success rate enhancement through intelligent reflecting surface optimization and dynamic resource allocation. Qi \textit{et al}.~\cite{10663259} presented an AoI-aware resource allocation scheme for RIS-assisted vehicular networks that achieves V2I link AoI reduction, V2V payload transmission success rate improvement, and spectral efficiency enhancement through DRL. However, in RIS-assisted communication scenarios, there exist complex coupling relationships among state variables. Existing DRL methods typically employ superficial state representations, relying mainly on simple feature concatenation and statistical correlations for decision-making. Such approaches tend to overlook the complex dependencies and intricate relationships inherent in the underlying state space.
%zw2:RIS?????????????????????????????????????????????????????????????????????????????????????????????????[1]??DL?RIS?????????????????????RIS??????????????????????????????????????????????????????????????[2]??DL?RIS?????????????????RIS????????????(SNR)???????6G???????(SE)?????(EE)???????????????????????[3]???????DL????RIS??????????(JoBNet)????????????????????????RIS???????????????????RIS?50%???????????????????????[4]???????????RIS???????????????PNAPG-Net???????????????????????????RIS???????????????????[5]??DL??????RIS?????DNNR-CRIS/DNNR,D-CRIS?????????????RIS??????????????????????????????????1/10?????IoT???????RIS?????????
%?????DL?RIS????????????????However???????????????????????????DL???????????????????????????????????????????????????????????????????????????
%???????DRL????????????????????????????????????RIS????????????????DL???DRL??????????????????????????????????????????????????????????DRL?RIS???????????????[7]???????????????(MA-DRL)??????????????????????RIS??????????????????????????????????????????[8]??MA-DRL??RIS?????D2D?????????????RIS???????????????????????????????????[9]??MA-DRL?RIS?????????????????????????????????????????????????????[10]??RIS??????AoI??????????DRL???V2I??AoI???V2V??????????????????However,?RIS???????????????????????????DRL?????????????????????????????????????????????????????

\section{Conclusion}
This paper introduces VariSAC, a GNN-augmented deep reinforcement learning framework designed to ensure time-continuous connectivity in RIS-assisted ISAC-enabled V2X networks. By formally defining the CCR and modeling resource allocation as a Markov Decision Process, VariSAC jointly optimizes channel allocation, power control, and RIS configuration for both communication and sensing, addressing the dual requirements of temporal continuity and probabilistic reliability. Extensive experiments on simulated and real-world urban datasets demonstrate that VariSAC consistently achieves superior performance in both V2I and V2V connectivity, delivering robust and persistent reliability under dynamic and resource-constrained scenarios. These results highlight the potential of VariSAC to enable reliable, continuous connectivity in next-generation vehicular networks.
%This paper addresses the challenge of establishing long-term reliable connectivity between vehicles in RIS-assisted ISAC-enabled V2X system. We formally define system reliability and model the problem as a Markov Decision Process, proposing the VariSAC framework as our solution. The framework employs a GNN-based deep learning approach to capture intrinsic relationships among vehicles, infrastructure, and RIS elements, while utilizing SAC for joint communication and sensing resource allocation to maintain system-wide reliable connectivity. Experimental results demonstrate that VariSAC significantly outperforms baseline methods across various resource-constrained scenarios using real-world road network data. Our solution not only ensures reliable system-level connectivity but also maintains sustained high-performance reliability for granular communication links, including both V2V and V2I connections.
%???????RIS??ISAC????????????????????????????????????????????????????????VariSAC??????????VariSAC????GNN???????????????????RIS???????????SAC?????????????????????????????VariSAC?????????????????????????????????????????????????????????????VariSAC??????????????????????????????????????????????????V2V???V2I?????????????????

%\section*{Acknowledgment}

%%\clearpage
\bibliographystyle{IEEEtran}
\bibliography{reference} 

\end{document}